\documentstyle[aas2pp4]{article}

\slugcomment{Accepted for publication in Ap.J.Letters., January 1997}
\lefthead{Belloni T. et al.}

\begin{document}

\def\spose#1{\hbox to 0pt{#1\hss}}
\def\lta{\mathrel{\spose{\lower 3pt\hbox{$\mathchar"218$}}
     \raise 2.0pt\hbox{$\mathchar"13C$}}}
\def\gta{\mathrel{\spose{\lower 3pt\hbox{$\mathchar"218$}}
     \raise 2.0pt\hbox{$\mathchar"13E$}}}
\def\Msun{{\rm M}_\odot}
\def\msun{{\rm M}_\odot}
\def\Rsun{{\rm R}_\odot}
\def\Lsun{{\rm L}_\odot}
\def\half{{1\over2}}
\def\RL{R_{\rm L}}
\def\zs{\zeta_{s}}
\def\zR{\zeta_{\rm R}}
\def\dJJ{{\dot J\over J}}
\def\dMM{{\dot M_2\over M_2}}
\def\tKH{t_{\rm KH}}
\def\eck#1{\left\lbrack #1 \right\rbrack}
\def\rund#1{\left( #1 \right)}
\def\wave#1{\left\lbrace #1 \right\rbrace}
\def\dd{{\rm d}}

\title{An unstable central disk in the superluminal black-hole
       X-ray binary GRS~1915+105}
\author{
        T.~Belloni\altaffilmark{1},
        M.~M\'endez\altaffilmark{1,2},
        A.R.~King\altaffilmark{3},
        M.~van der Klis\altaffilmark{1},
        J.~van Paradijs\altaffilmark{1,4}
       }

\altaffiltext{1} {Astronomical Institute ``Anton Pannekoek'',
       University of Amsterdam and Center for High-Energy Astrophysics,
       Kruislaan 403, NL-1098 SJ Amsterdam, the Netherlands}

\altaffiltext{2}{Facultad de Ciencias Astron\'omicas y Geof\'{\i}sicas, 
       Universidad Nacional de La Plata, Paseo del Bosque S/N, 
       1900 La Plata, Argentina}

\altaffiltext{3}{Astronomy Group, University of Leicester,
                 Leicester LE1 7RH, United Kingdom}

\altaffiltext{4}{Physics Department, University of Alabama in Huntsville,
       Huntsville, AL 35899, USA}

\begin{abstract}

We have analyzed the X-ray spectra of the microquasar GRS~1915+105, as 
observed with the PCA on the Rossi XTE, during periods of stable weak 
emission, outbursts and rapid flaring. We find that the complicated X-ray 
intensity curve of this source can be described by the rapid removal and
replenishment of matter forming the inner part of an optically thick accretion
disk, probably caused by a thermal-viscous instability analogous to that
operating in dwarf novae, but here driven by the Lightman--Eardley
instability. We find that the mass accretion rate in quiescence is 
$\sim 10^{-6}$ M$_{\odot}$~yr$^{-1}$. Only a small fraction of the energy
liberated by accretion is emitted as radiation. We suggest that most of
this energy is advected into the black hole in the high--viscosity state of
the outburst cycle.

\end{abstract}

\keywords{accretion, accretion disks ---
          binaries: close --- black hole physics -- instabilities
          -- X-rays: stars --- stars: individual GRS~1915+105}

\section{INTRODUCTION}

The X--ray source GRS~1915+105 was the first galactic object to show
superluminal expansion in radio observations (\cite{mr94}). The standard
interpretation of this phenomenon in terms of relativistic jets
(\cite{res66}) unambiguously places the source at a distance $D = 12.5$ kpc,
with the jet axis at an angle $i = 70^\circ$ to the line of sight
(\cite{mr94}). GRS~1915+105 was discovered as an X-ray transient in 1992
(\cite{cas92}). It was found to be in a bright state with the Rossi X-ray
Timing Explorer (RXTE) in April 1996, and has been bright in X rays since.
During this period it has shown a remarkable richness in its variability
behavior. It showed quasi-periodic oscillations, with frequencies ranging
from $10^{-3}$ Hz to 67 Hz (\cite{mrg96}). On several occasions its
light curve showed a complicated pattern of dips and rapid transitions
between high and low intensity, which repeated on a time scale between 30
and 400 minutes (\cite{gmr96}). The source is suspected to be a black hole
binary because of its similarities to the other galactic superluminal source
GRO~J1655-40 (\cite{zh94}), which has a dynamical mass estimate 
implying a black
hole (\cite{bai95}), and because it has often been well above the Eddington
luminosity for a neutron star. Here we show that the complicated light curve
of GRS~1915+105 can be described by the rapid ($\sim$1 s) appearance and
disappearance of emission from an optically thick inner accretion disk,
probably caused by a thermal-viscous instability analogous to that operating
in dwarf novae. 

\section{X--RAY OBSERVATIONS}

The observations discussed here were carried out with the PCA on the
Rossi XTE, on 1996 June 19, September 27 and October 1 to 29.
During these observations the source was both very luminous ($L_{\rm X}
\simeq 5 \times 10^{38}$ -- $3 \times 10^{39}$ erg s$^{-1}$ at $D = 12.5$
kpc in the 2--50 keV band) and very variable (see Fig. 1), similar to
the observations reported by Greiner, Morgan, \& Remillard (1996) and
Morgan, Remillard, \& Greiner (1997).

The variability is highly complex, but quite structured. We identify three
types of variability (``states'') from the light curves (see Fig. 1):
a {\it quiescent} state at a relatively low count rate and with little
variability, an {\it outburst} state at high count rate with strong
red-noise variability, and a {\it flare} state where the flux shows rapid
alternations between two flux levels which can be extremely 
complex in shape. When more than one state is observed within the same
observation (a typical observation spans a few hours), the sequence is
always the same: quiescent-outburst-flare-quiescent etc., sometimes with a
highly repetitive pattern. In some observations, the source is observed in
only the quiescent or flaring state, but never only in the outburst state.
These states are similar to the lull, flare and sputter state described
in Greiner, Morgan \& Remillard (1996) on the basis of previous observations
of the source. 

In this Letter we take the observation of 1996 October 7th (6:36 to 7:36 UT;
Fig. 1) as a prototype containing all three states. The highly--organized
complex variability also shows up in a plot of hardness ratio versus time
(Fig. 1, bottom panel). The count rate and hardness ratio histories can be
combined in a hardness-intensity diagram (Fig. 2). The approximately
hyperbolic shape of this plot suggests that to a first approximation,
the source can be understood as the superposition of a constant and a varying
source, each with nearly constant hardness ratio. An explicit example
of this is seen in the data of Fig. 1 (bottom panel), where the hardness 
ratio varies in a near square--wave pattern as the count rate alternates 
between a low and high value. The spectra in these two brightness levels 
are strikingly different (inset frame in Fig. 2), mainly because of a large
excess of emission below 20 keV at high count rate. This would appear to be
the varying component invoked above. We confirmed this idea by the following
procedure.

We accumulated X-ray spectra for the quiescent and outburst intervals
for the observation of 1996 October 7th (Fig. 2, inset). Fits with simple
single-component models did not give acceptable results. We obtained good
fits (reduced $\chi^2$=1.2 and 0.8 respectively) with a ``standard'' model
for black-hole candidates, consisting of a multi-temperature disk-blackbody
plus a power law (\cite{mal84}). The addition of an iron line plus
absorption edge was necessary to fit the data around 6 keV. All fits were
performed in the range 2 -- 50~keV, and a 1\% systematic uncertainty was
added to the data to account for the uncertainties in the response matrix
calibration (\cite{cui96}). The best fit parameters, using the known
distance and inclination (we assume that the latter is the same as the
inclination of the radio jets to the line of sight), are shown in Table 1. 
For the outburst state we find a column density 
N$_{\rm H}$ = 8.0$\times 10^{22}$ cm$^{-2}$, somewhat higher than the
ROSAT value (6$\times 10^{22}$ cm$^{-2}$, see Greiner, Morgan, \& Remillard
1996). For the quiescent state we have fixed N$_{\rm H}$ to the outburst
value. The marked spectral changes between quiescence and outburst are due
to a steepening of the power law component and a large change in temperature
of the disk-blackbody. From additional fits to other parts of the light
curve and to other observations we find that while the power law index can
assume values anywhere in the range 2.3--3.9, the inner temperature of the
disk-blackbody component switches between two values: $\sim$0.5 keV, which
corresponds to an inner radius of several hundred kilometers, and $\sim$2.2
keV, with an inner radius of $\sim$20 km. This hot disk component switches
on at count rates larger than 20000 c/s (see Fig. 2), even during the brief
peaks of the oscillating events in the flaring state. Below 20000 c/s, the
hardness ratio changes in Fig. 2 are due primarily to changes in the
power-law index. 

By looking at the light curve in Figure 1 at high time resolution, one sees
that the fastest rise and fall times for the flare state remain constant.
The shortest doubling rise times are $\sim$2 s, while the corresponding
shortest decay times are significantly less, i.e. 0.5 -- 1.0 s. Two further
characteristic time scales apparent in the X--ray light curve (cf. Fig. 1)
are the duration of the quiescent state and of a complete
quiescent-outburst-flare cycle. For the light curve plotted in Fig.~1, these
are $\sim$300 s and $\sim 1000$ s, respectively.

\section{INTERPRETATION}

If we take the simplest possible interpretation of the multi--temperature
blackbodies, i.e. as the emission from optically thick regions of an
accretion disk around a compact object of mass $M$, 
we have $R_{\rm in,\ v} \simeq 20$ km
for the brightest state of the varying source, and $R_{\rm in,\ c}\sim$300
km for the constant source. From $R_{\rm in}$ and $T_{\rm in}$ we can
infer the mass accretion rate according to 
\begin{equation}
\dot M = 8 \pi R_{\rm in}^3 \sigma T_{\rm in}^4 / 3 GM.
\label{e0}
\end{equation}
$R_{\rm in, v}$ cannot be smaller than the innermost stable orbit
around a black hole. This
gives $M \lta 2.4\msun, \dot M \gta 1\times 10^{-7}$ $\msun$ yr$^{-1}$ 
in outburst and $\dot M \gta 6.8\times 10^{-7}$ $\msun$ yr$^{-1}$
in quiescence for a Schwarzschild hole ($R_{\rm in, v} \geq
6GM/c^2$). For an extreme Kerr hole we have $R_{\rm in, v} \geq GM/c^2$,
so $M < 14 \msun$, and $\dot M \gta 1.5\times 10^{-8}$ $\msun$ yr$^{-1}$
in outburst and $\dot M \gta 1\times 10^{-7}$ $\msun$ yr$^{-1}$ in quiescence.
The fact that the inferred accretion rates anticorrelate with the total
luminosity implies that the accretion yield is lower when the accretion rate
is high. This can be accomplished by trapping of the radiation and
quasi--spherical infall (``advection'') of the accretion energy into
a black hole
(cf \cite{kat77}, \cite{beg78}, \cite{nar96}). We shall see that the disk
interpretation indeed suggests that just this occurs.

We arrive at the following picture. The ``constant''
source is the emission from the outer parts of the accretion disk. This
component varies only on the long viscous time scales characteristic
of the outer disk. The varying source is the emission from the inner disk
region and from a hot comptonizing corona above the disk, whose spectrum
varies in  response to changes in the disk continuum. As we shall explain
below, the inner disk region empties and refills on time scales of seconds,
thus causing the rapid flux and spectral changes. Radiation pressure becomes
important in the inner region: it exceeds gas pressure for disk radii
\begin{equation}
R \lta 10^8\alpha^{0.1}\dot M_{18}^{0.8}M_{10}^{-0.14}\ {\rm cm} 
\label{e1}
\end{equation}
(cf \cite{fkr92}). Here $\alpha$ is the Shakura--Sunyaev viscosity
parameter, $\dot M_{18}$ is the accretion rate in units of 
$10^{18}$ g s$^{-1}$, and $M_{10}$ is the black hole mass in units of 
$10\msun$. In such regions the disk scale height $H$ is almost constant
with radius, so that
\begin{equation}
{H\over R} \simeq 0.15\dot M_{18}R_7^{-1}
\label{e2}
\end{equation}
(cf \cite{fkr92}) and the thin disk approximation necessarily
breaks down near the inner edge of the disk ($R_7 \sim 0.1$). Here $R_7$
is the radial coordinate in units of $10^7$ cm.

Thermal balance at a given radius of a disk can be expressed in terms of
a relation between the local accretion rate $\dot M$ and the surface
density $\Sigma$. It is well known that for regions with $P_{\rm rad}
\sim P_{\rm gas}$, the $\dot M(\Sigma)$ curve has a branch with a negative
slope, which is thus thermally and viscously unstable (\cite{le74}).
As pointed out by Abramowicz et al. (1988), Lasota \& Pelat (1991) and
Chen et al. (1995), inward advection of the hot gas in the high $\dot M$
state is likely to stabilize the system. This leads to an S--shaped thermal
balance curve in the $\dot M - \Sigma$ plane, at typical disk effective
temperatures $\sim$ few $\times 10^6$ K on the lower branch. This is very
similar to the familiar thermal balance curve found near the hydrogen
ionization temperature $\sim 10^4$ K in dwarf nova disks. In the same way
as there, the disk is locally thermally stable on the branches with positive
slope, and locally unstable on the middle branch with negative slope. We
obviously again have the possibility of limit--cycle behavior, this time
only in central regions of the disk with temperatures exceeding a few
million degrees. Cannizzo (1996) has modeled the variability of the
bursting pulsar GRO~J1744--28 (\cite{kou96}) in these terms. For that source 
the large assumed value of the inner disk radius (because of the neutron
star magnetic field) leads to much longer time scales than for GRS~1915+105. 
We note that the Lightman-Eardley instability is only expected to operate
inside the radius given by Eq. (\ref{e1}), which indeed is close to the inner
radius we estimate for the ``constant'' source.

Since the upper branch of the S--curve involves physics which is currently
unclear, we confine ourselves to estimating the time scales of the X--ray
variation. After an outburst, the outer disk attempts to refill the central
disk via steady inflow on its viscous time scale, initially causing the slow
recovery during the quiescent phase. A brightness rise occurs as a heating
wave propagates outwards into the central disk from its inner edge before
it has been refilled to a steady--state density profile. The hot radiating 
matter is removed by infall into the central black hole on the effective
viscous time in the outbursting state. The latter time is not well known, as 
the thin disk approximation fails in this state. Heating fronts and infall
can alternate many times in the flaring state, before the central disk
eventually returns to the quiescent state. The whole cycle repeats roughly
on the viscous time of the outer disk.

We can estimate the expected time scales as follows. The rise time is the 
time for the heating wave to pass through the central disk. Assuming that the 
heating front travels with subsonic velocity, the sound speed $c_S$ being 
calculated in the cool ($\lta 6\times 10^6$ K) disk state, we get
\begin{equation}
t_{\rm rise} \gta {R_{\rm out}\over c_S} \simeq 1 R_{\rm out,\ 7}\ {\rm s}.
\label{e3}
\end{equation}
where $R_{\rm out,\ 7} = R_{\rm out}/10^7$ cm.
The decay time cannot be shorter than the free--fall time scale, giving
\begin{equation}
t_{\rm decay} \gta 0.03 R_{\rm out,\ 7}^{3/2}M_{10}^{-1/2}\ {\rm s}.
\label{e4}
\end{equation}
The whole cycle should repeat roughly on the viscous time scale at the
outer edge of the central disk, i.e.
\begin{equation}
t_{\rm rec} = {R^2\over \nu} = {R\over H}{R\over \alpha c_S} = 
     6\alpha^{-1}\dot M_{18}^{-1}R_7^2\ {\rm s},
\label{e5}
\end{equation}
where we have written the kinematic viscosity as $\nu = \alpha c_sH$ and used
(Eqs. \ref{e2} and \ref{e3}). 
We note that (Eqs. \ref{e3} and \ref{e4}) are in rough agreement
with the observed rise and decay times $\sim 2$ s, $\sim 0.5$ s, while 
typical values $\alpha \sim 10^{-2}$ assumed for quiescent disks imply
recurrence times $\sim 600R_7^2$ s, quite close to the observed values.
Note that the X-ray rise and fall episodes differ from the case considered by
Cannizzo (1996). There the infall on to the central neutron star is the
process which raises the X-ray production rate, whereas here infall causes
hot matter to be lost down the black hole, leading 
eventually to a decay of the X-rays.

We note that the several features of the X-ray light curves strongly
resemble the results of dwarf nova disk instability calculations.
Even though the basic cause of the instability differs here, the general
picture of heating and cooling fronts travelling back and forth leads us
to expect generic similarities. In particular, alternating wide and narrow
outbursts, and series of narrow outbursts followed by a wide one are 
common features (see \cite{jvp83}, for the outburst width distribution for
dwarf novae).
These typically result when the heating wave stalls before
crossing the entire unstable region. Similarly, a generic prediction
of this type of model comes from the fact that the rise and decay
episodes are fundamentally asymmetric. When observed with high enough
time resolution, the system should perform a hysteresis loop in an X-ray
color-color diagram. Unfortunately, the large variations in the power
law component do not currently allow one to draw a conclusion from simple
color-color plots. The accumulation of high time resolution energy
spectra, not available in the current data, will allow one to test for
the presence of such a hysteresis loop. A full quantitative discussion will
be presented in a future paper.

The extremely high values of the mass accretion rate we infer from the
spectral fits in the quiescent state do not lead to a correspondingly high
X-ray luminosity generation in the inner disk (with a canonical efficiency
of 0.2 c$^2$ per gram one would have expected an X-ray luminosity in excess
of 10$^{40}$ erg/s). This indicates that a major fraction of the 
gravitational energy
released in the inner disk is not emitted as radiation, suggestive of
advection into a black hole. Assuming that the accretion
rates derived from our spectral model are essentially correct and
that ejection is relatively unimportant, a substantial
fraction of the inflow disappears into the central object without broadcasting
a signal to the outside world. This supports the idea that it does not have a 
solid surface, and thus has the most characteristic property of a black hole.
We note that the interpretation proposed in this paper depends on the
spectral model we have used. We emphasize that at this point it is
difficult to exclude absolutely the possibility that the compact object in
GRS~1915+105 is a neutron star, as the study of X-ray spectra of neutron
stars accreting at a hyper--Eddington rate is fairly unexplored.
However, a black hole surrounded by an
optically thick accretion disk of variable inner radius provides a natural
explanation for the otherwise apparently intractable phenomenology
presented by GRS~1915+105.

It is improbable that the high inferred accretion rate represents the
high state of some much longer-term instability in the outer disk.
It seems more likely that the mass transfer rate is always very high, and
that the outbursts of GRS~1915+105 are not caused by an increase of the
mass transfer rate, but quite possibly by a decrease from a value high
enough to smother the X-ray emission. In the smothered state, the object
would then appear only as a bright infrared source. Average mass transfer
rates of the required order ($10^{-7}$--$10^{-6}\msun$/yr) can easily be
supplied for example by a donor star crossing the Hertzsprung gap, and
thus undergoing thermal--timescale radius expansion (\cite{kkf96}).

\acknowledgements

MM is a fellow of the Consejo Nacional de Investigaciones
Cient\'{\i}ficas y T\'ecnicas de la Rep\'ublica Argentina.
This work was supported in part by the Netherlands Organization for
Scientific Research (NWO) under grant PGS 78-277 and the Netherlands
Foundation for Research in Astronomy (ASTRON) under grant 781-76-017.
ARK gratefully acknowledges a PPARC Senior Fellowship and the hospitality
of the Astronomical Institute ``A. Pannekoek'' at UvA. JVP acknowledges
support from NASA under contract NAG 5-3003. We thank Keith Jahoda for
helping with the calibration of the PCA data, and the RXTE team for making
these data publicly available.

{}

\clearpage

\figcaption[]{Upper panel: 2.0--13.3 keV
             PCA light curve.
             Time zero corresponds to 1996 October 7th 6:36 UT.
             Lower panel: corresponding hardness ratio (13.3--60.0 kev /
             2.0--13.3 keV).
             Examples of the three states can be identified 
             as follows. Quiescent
             state: T$\sim$400--800; outburst state: T$\sim$800--1200;
             flaring state: T$\sim$1200--1700.
            }

\figcaption[]{Hardness-intensity diagram for the data in Fig. 1.
              Rate and hardness ratio are defined as in Fig. 1.
              Inset: PCA energy spectra for the average quiescent 
              (dots) and outburst (circles) states.
            }

\clearpage

\begin{deluxetable}{lcccccc}
\footnotesize
\tablecaption{
Best-fit parameters for the quiescence and outburst
spectra of October 7th. N$_{\rm H}$ is the column absorption in 10$^{22}$
cm$^{-2}$,
kT$_{\rm in}$ is the inner temperature of the
disk-blackbody (in keV), R$_{\rm in}$ is the inner radius in km,
$\Gamma$ is the power-law photon index. The (unabsorbed) fluxes are in
units of erg cm$^{-2}$ s$^{-1}$ in the 2--50 keV band. \label{tbl-1}}
\tablewidth{0pt}
\tablehead{
\colhead{State}     &
\colhead{N$_{\rm H}$}   &
\colhead{kT$_{in}$}   &
\colhead{R$_{in}$}  &
\colhead{$\Gamma$}  &
\colhead{DBB Flux}  &
\colhead{PL flux}
}
\startdata
Quiescence  &  8.0                             & 0.57$\pm$0.03  &
               319$\pm$9                       & 2.22$\pm$0.05  &
               (0.96$\pm$0.01)$\times 10^{-8}$ &
               (0.80$\pm$0.03)$\times 10^{-8}$ \nl

Outburst    &  8.0$\pm$0.5                     & 2.27$\pm$0.03  &
               20.3$\pm$0.3                    & 3.57$\pm$0.03  &
               (3.99$\pm$0.07)$\times 10^{-8}$ &
               (4.65$\pm$0.22)$\times 10^{-8}$ \nl

\enddata
\end{deluxetable}

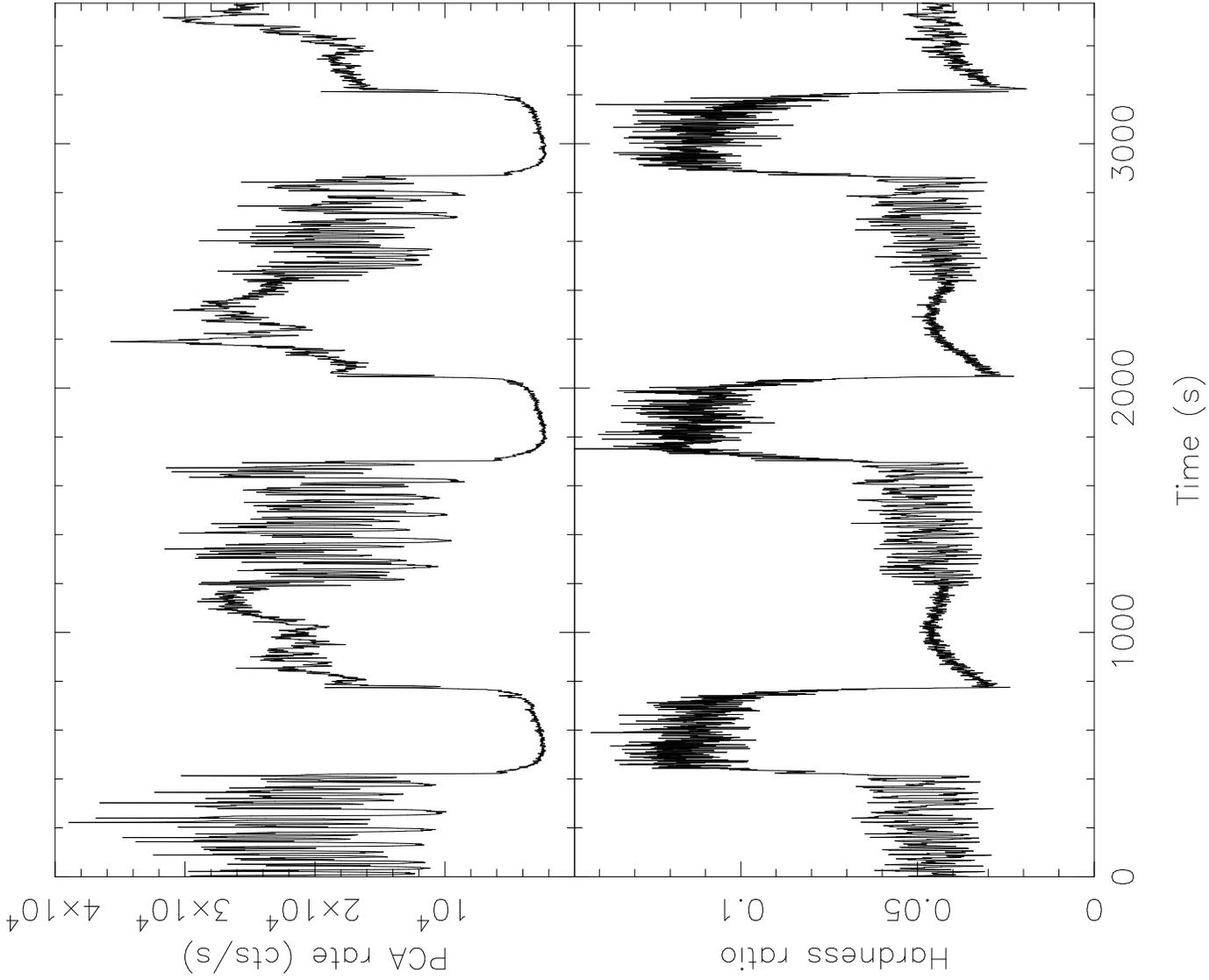
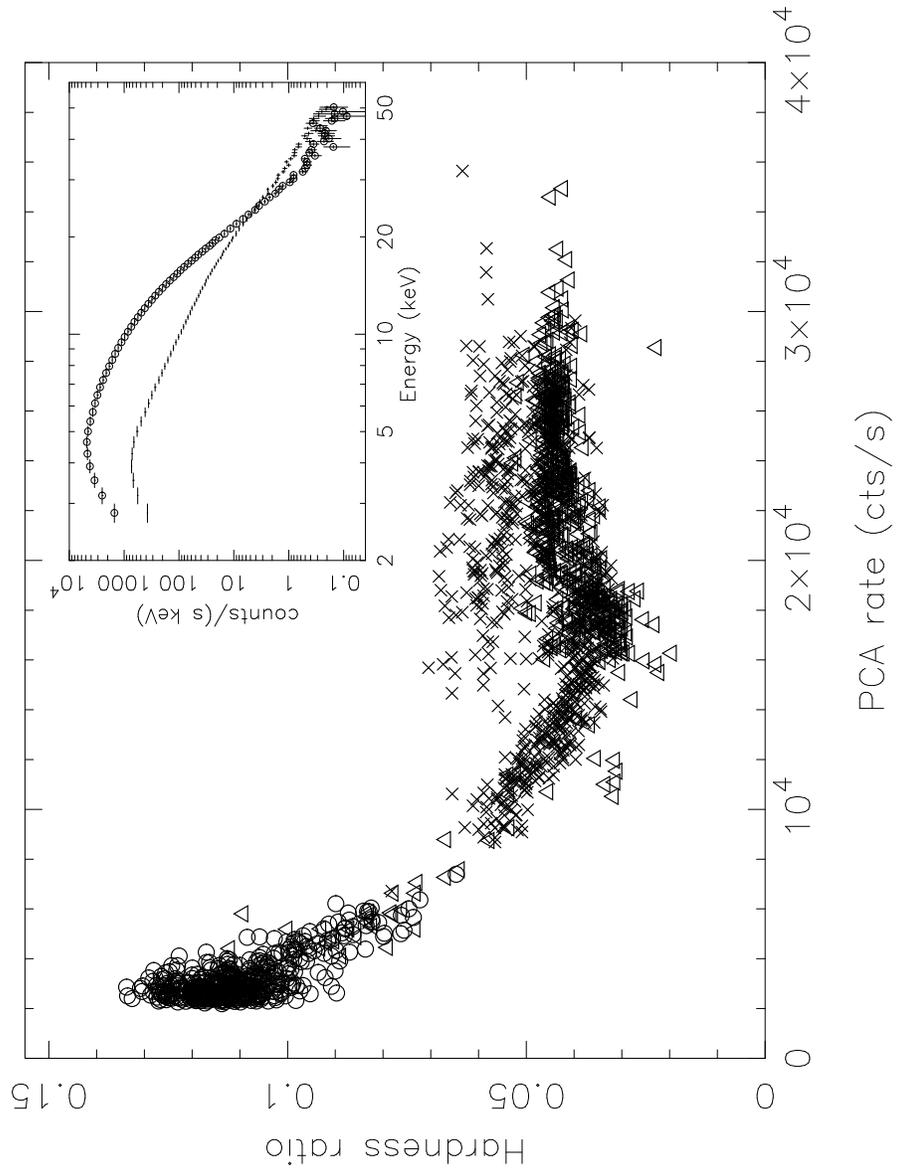
\end{document}